\documentclass[]{revtex4}

\usepackage{graphicx}

\newcommand{\ket}[1]{\, | #1 \rangle}
\newcommand{\text}[1]{\mbox{#1}}
\newcommand{\texts}[1]{\mbox{\scriptsize #1}}
\newcommand{\om}{\omega}
\newcommand{\Om}{\Omega}
\newcommand{\la}{\lambda}
\newcommand{\ga}{\gamma}

\newcommand{\De}{\Delta}
\newcommand{\de}{\delta}

\begin{document}

\title{Self-induced transparency and giant nonlinearity 
in doped photonic crystals}

\author{Gershon Kurizki}
 \email{gershon.kurizki@weizmann.ac.il}
 \homepage{http://www.weizmann.ac.il/chemphys/gershon/}
\author{David Petrosyan}
\affiliation{Department of Chemical Physics, Weizmann Institute of Science, 
Rehovot 76100, Israel}

\author{Tomas Opatrny}
\affiliation{Department of Physics, Texas A\&M University,\\
College Station, Texas 77843-4242}

\author{Miriam Blaauboer}
\affiliation{Lyman Laboratory of Physics, Harvard University, Cambridge, 
Massachusetts 02138}

\author{Boris Malomed}
\affiliation{Department of Interdisciplinary Studies, Faculty of
Engineering, Tel Aviv University, Tel Aviv 69978, Israel}

\date{\today}

\begin{abstract}
Photonic crystals doped with resonant atoms allow for uniquely advantageous 
nonlinear modes of optical propagation: (a) Self-induced transparency
(SIT) solitons and multi-dimensional localized "bullets" propagating at 
photonic band gap frequencies. These modes can exist even at ultraweak 
intensities (few photons) and therefore differ substantially either from
solitons in Kerr-nonlinear photonic crystals or from SIT solitons in
uniform media. (b) Cross-coupling between pulses exhibiting electromagnetically
induced transparency (EIT) and SIT gap solitons. We show that extremely strong 
correlations (giant cross-phase modulation) can be formed between the two 
pulses. These features may find applications in high-fidelity classical 
and quantum optical communications.
\end{abstract}

\pacs{Keywords: Coherent optical effects; pulse propagation and solitons;
Kerr effect}

\maketitle

\section{Introduction} 

Photonic crystals (PCs) can exhibit an interplay between Bragg reflections, 
which block the propagation of light in photonic band gaps (PBGs)
\cite{pbg-bib,yablon,john,joannopoulos,defect_a,defect_b}, and the 
dynamical modifications of these reflections by {\em nonlinear} light-matter
interactions \cite{kofman,john_wang,pl,Ze95,Scal94,Scal94a}. A very 
interesting situation arises when foreign atoms or ions---dopants---with 
transition frequencies within the PBG are implanted in the PC 
\cite{kofman,john_wang,pl}. Then light near one of these frequencies 
resonantly interacts with the dopants and is concurrently affected by the 
PBG dispersion. Consequently, highly nonlinear processes with a rich variety 
of unusual PC-related features are anticipated. 

Our aim in recent years has been to identify those regimes of nonlinear
optical propagation in doped PCs that allow transmission of extremely weak 
pulses, while filtering out undesirable noise, and are therefore highly 
advantageous for optical communications, data storage and processing, near 
or at the quantum limit. These requirements are satisfied by novel regimes 
surveyed in this paper that have been theoretically discovered and 
investigated by us:
a) {\em Self-induced transparency} (SIT) solitons propagating inside or near
a PBG at a frequency that is near resonant with the transition frequency of
the dopant \cite{Kuri2001,Kozh95,Kozh98a,Opat99}. This peculiar form of gap 
solitons (GSs) is immune to resonant absorption even for a {\em small number 
of photons}, and may also possess two- or three-dimensional (2D or 3D) 
localization in the form of light bullets (LBs)\cite{Blaa00a} 
(Sec. {\ref{sit_rabr}}).
b) {\em Cross-coupling of SIT and electromagnetically-induced transparency 
(EIT) pulses} in PCs. We put forward a new regime in Sec. {\ref{sec:eit_sit}}: 
a strong modulation of the phase of a weak pulse subject to EIT by a control 
pulse in the form of an SIT GS moving at the {\em same} slow velocity. Thereby 
giant cross-phase modulation can be formed between the GS and the EIT pulses.
In Sec. {\ref{sec:concl}} we summarize and discuss our findings in this paper.

\section{Self-induced transparency (SIT) gap solitons\label{sit_rabr}}

\subsection{Background}

A GS is usually understood as a self-localized moving or standing
(quiescent) bright region, where light is confined by Bragg
reflections against a dark background.  The soliton spectrum is tuned
away from the Bragg resonance by the nonlinearity at sufficiently high
field intensities. The first type of GS had been predicted 
\cite{Chri89,Acev89,Feng93,Ster94}, and later observed \cite{Eggl96},  
in a Bragg grating possessing Kerr-nonlinearity. A principally different 
mechanism of GS formation has been theoretically discovered by our group 
in a periodic array of thin layers of {\em resonant two-level atoms} (TLA) 
separated by half-wavelength nonabsorbing dielectric layers, i.e., a 
{\em resonantly absorbing Bragg reflector} (RABR) 
\cite{Kuri2001,Kozh95,Kozh98a,Opat99}. As opposed to the $2\pi$-solitons 
arising in SIT, i.e., resonant field--TLA interaction in uniform media 
\cite{McCa67,McCa69}, their GS counterparts in a RABR may have an 
{\em arbitrary} pulse area \cite{Kozh95,Kozh98a}. It must be stressed that
stable, moving or standing, GS solutions have been consistently obtained only 
in a RABR with {\em thin} active TLA layers. By contrast, the case of a 
periodic structure {\em uniformly} doped with active TLA calls either for a 
solution of the wave (Maxwell) equation {\em without} the spatial 
slowly-varying envelope approximation (SVEA), or for a solution of an 
{\em infinite} set of coupled Bloch equations for all spatial harmonics of 
the atomic polarization (Fourier components) \cite{Kuri2001}. Therefore, 
an attempt \cite{Akoz98} to obtain a self-consistent solution for a 
uniformly-doped periodic structure by imposing the SVEA, or by arbitrarily 
truncating the infinite hierarchy of equations for the harmonics of atomic
population inversion and polarization to its first two orders, is generally
unjustified. In fact, it can be shown numerically to fail for many parameter 
values.

In the simplest case of a uniform (bulk) medium, when the driving field is 
resonant with the atomic transition, the TLA Bloch equations can be easily 
integrated and the Maxwell equation then reduces to the sine-Gordon equation
\begin{equation}
\frac{\partial^2 \theta}{\partial \zeta \partial \tilde{\tau}} = -\sin{\theta}
\label{sn_gr}
\end{equation}
for the pulse area $\theta=\int_{-\infty}^{t} \Om \, d t'$, i.e., the time
integral of the Rabi frequency $\Om$. Equation (\ref{sn_gr}) is written in 
terms of the dimensionless variables $\tilde{\tau}=(t-n_0 z /c)/\tau_0$ and 
$\zeta=n_0 z/c\tau_0$, where 
$\tau_0= \frac{n_0}{\mu}\sqrt{\frac{\hbar }{2\pi \omega_{c} \varrho_{0}} }$, 
is the cooperative resonant absorption time, $\varrho_{0}$ being the TLA 
density (averaged over $z$), $\mu$ the dipole moment of the TLA transition at 
frequency $\om_0$ and $n_0$ is the refraction index of the host medium.
This sine-Gordon equation is known to have solitary-wave solutions, which 
propagate without attenuation or distortion with a conserved pulse area of
$2\pi$ \cite{McCa67,McCa69}. These SIT solitons have the form:
\begin{equation}
{\Om}(\zeta, \tilde{\tau}) = 
(\tau_0)^{-1} A_0 \text{sech} [ \beta (\zeta-v \tilde{\tau})] , \label{2pi-SG}
\end{equation}
where the pulse width $\beta$ is an arbitrary real parameter uniquely defining 
the amplitude $A_0=2/\beta$ and group velocity $v=1/\beta^2$ of the soliton.
In what follows, Eq. (\ref{2pi-SG}) will be compared with an SIT GS in a RABR.

\subsection{SIT in RABR: The Model 
\label{S:model}}

\begin{figure}[t]
\centerline{\includegraphics[height=5cm]{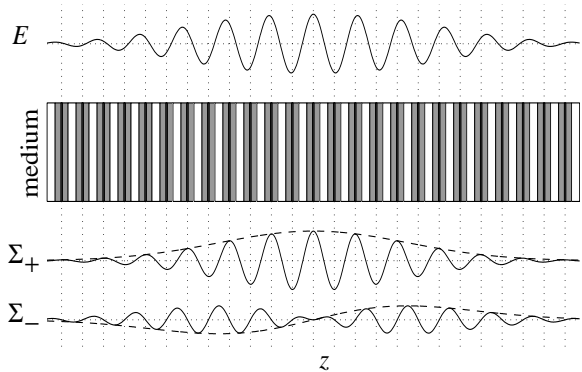}}
\caption{Schematic description of the periodic RABR and of the decomposition 
of the electric field $E$ into modes $\Sigma_{+}$ and $\Sigma_{-}$.
The shading represents regions with different index of refraction; the darker
the shading the larger $n$ is. The black regions correspond to the TLA layers.
The localization of the field envelope over $\sim 20$ structure periods
is shown for the sake of visualization; in reality, the field is localized
over a hundred or more periods.
\label{figschem}}
\end{figure}

Let us assume \cite{Kuri2001,Kozh95,Kozh98a,Opat99} a one-dimensional (1D)
periodic modulation of the linear refractive index 
$n^2(z)=n_0^2[1+a_{1}\cos (2k_{c}z)]$. The periodic grating gives rise to 
a PBG with a central frequency $\om_{c} =k_{c}c/n_0$ and gap edges at 
$\omega _{1,2}=\omega_{c} \left( 1\pm a_1/4\right)$. The electric field 
$E$ of a pulse propagating along $z$ can be expressed by means of the 
dimensionless quantities 
$\Sigma_{\pm} \equiv 2 \tau_o \mu \hbar^{-1} 
({\cal E}_F \pm {\cal E}_B )$, where
${\cal E}_F$ and ${\cal E}_B$ denote the slowly varying amplitudes of the
forward and backward propagating fields, respectively, as
\begin{equation}
E(z,t)=\hbar (\mu \tau _0)^{-1}
\left\{ {\rm Re} [ \Sigma_{+}(z,t)e^{-i\om_{c} t} ] \cos k_{c}z  
  -{\rm Im} [ \Sigma_{-}(z,t)e^{-i\om_{c} t} ] \sin k_{c}z \right\} 
. \label{be4}
\end{equation}
We further assume that {\em very thin} TLA layers (much thinner than 
$1/k_{c}$), whose resonance frequency $\om_0$ is close to the gap center
$\om_c$, are placed at the maxima of the modulated refraction
index (Fig. \ref{figschem}). They are located at positions $z_{j}$ such that
\begin{equation}
 e^{i k_{c}z_{2j}} = 1, \qquad e^{i k_{c}z_{2j+1}} = -1 ,
\label{expons}
\end{equation}
i.e., the TLA density is described by 
$\varrho = \varrho_{0} \la/2 \sum_{j} \de(z-z_{j})$, where $\la = 2 \pi/k_c$
is the wavelength. 

The Bloch equations for the slowly varying polarization envelope $P$ and 
inversion $w$ in the even numbered layers can be obtained (in the slowly
varying envelope approximation) by substituting for the Rabi frequency
$\Om = \tau_{0}^{-1} (\Sigma_+ \cos k_{c}z + i \Sigma_- \sin k_{c}z )$
and applying Eq. (\ref{expons})  at the positions of these layers:
\begin{eqnarray}  
  \frac{\partial P}{\partial \tau } &=&-i\delta P+\Sigma _{+}w,  \label{P} \\
  \frac{\partial w}{\partial \tau } &=&-{\rm Re}\ \left( \Sigma
  _{+}P^{*}\right) .  \label{w}
\end{eqnarray}
Combining Eqs. (\ref{P}) and (\ref{w}), one can eliminate the TLA population 
inversion: $w= \sqrt{1-|P|^2}$. The remaining equation, together with the 
Maxwell equations for $\Sigma_{\pm}$ (driven by $P$), form a {\em closed 
system},
\begin{eqnarray}
\frac{\partial ^2\Sigma _{+}}{\partial \tau ^2}-\frac{\partial ^2\Sigma _{+}}
{\partial \zeta ^2} &=& \eta ^2\Sigma _{+}+2i(\eta -\delta )P
- 2\sqrt{1-|P|^2} \Sigma_{+},  \label{bef1} \\
\frac{\partial ^2\Sigma _{-}}{\partial \tau ^2}-\frac{\partial ^2\Sigma _{-}}
{\partial \zeta ^2}&=&-\eta ^2\Sigma _{-}-2\frac{\partial P}{\partial \zeta}
, \label{driven} \\
\frac{\partial P}{\partial \tau } &=& - i\de P-\sqrt{1-|P|^2} \Sigma_{+},  
\label{bef2} 
\end{eqnarray}
where $\tau \equiv t/\tau_0$, $\zeta \equiv \left( n_0/c\tau _0\right) z$
and $\de \equiv (\om_0-\om_c )\tau_0$ are the dimensionless time, coordinate, 
and detuning, respectively, and 
$\eta = l_{\rm abs}/l_{\rm refl} = a_{1}\om_c \tau_0/4$ is the 
dimensionless modulation strength, which can be expressed as the ratio of the
TLA {\em absorption distance} $l_{\rm abs}=\tau_0 c/n_0$ to the {\em Bragg 
reflection distance\/} $l_{\rm refl}=4c/(a_1 \omega_c n_0)$. We emphasize
that the above equations are obtained using the SVEA, which is valid under 
the assumption that the Bragg reflection does not appreciably change the 
pulse envelope over a distance of a wavelength, $l_{\rm refl} \gg \la$, 
whence $a_1 \ll 2/\pi$.

To reach general understanding of the dynamics of the model, one should first 
consider the spectrum produced by the {\em linearized} version 
of Eqs. (\ref{bef1})--(\ref{bef2}), which describes {\em weak fields} in the 
limit of infinitely thin TLA layers.  Setting
$\Sigma_+=A e^{i(\kappa \zeta -\chi \tau )}$, 
$\Sigma_-=B e^{i(\kappa \zeta -\chi \tau )}$, 
$w=-1$, and $P = C e^{i(\kappa \zeta -\chi \tau )}$, 
we obtain from the linearized equation (\ref{bef2}) that 
$C=i(\delta -\chi )^{-1}A$.
Substituting this into Eqs. (\ref{bef1}) and (\ref{driven}), we arrive at the
dispersion relation for the wavenumber $\kappa $ and frequency $\chi $ in
the form 
\begin{equation}
(\chi^2-\kappa^2-\eta^2)(\chi -\de ) \times \left\{ (\chi -\de )
\left[ \chi^2 - \kappa^2 - (2 + \eta^2)\right] +2(\eta -\de ) \right\} = 0.  
\label{dispersion}
\end{equation}

\begin{figure}[t]
\centerline{\includegraphics[height=6cm]{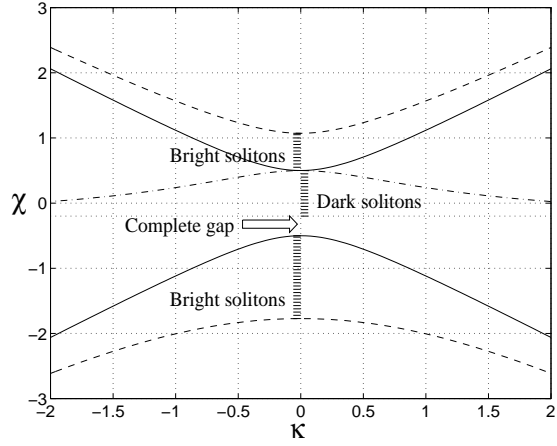}}
\caption{The RABR dispersion curves at $\eta =0.5$ and $\delta =-0.2$. 
The solid lines show the dispersion branches corresponding to the `bare'
(noninteracting) grating, while the dashed and dash-dotted lines
stand for the dispersion branches of the grating `dressed' by the
active medium. The frequency bands that support the standing dark and
bright solitons are shaded.  
\label{disp}}
\end{figure}

Different branches of the dispersion relation generated by 
Eq. (\ref{dispersion}) are shown in Fig.~\ref{disp}.  The roots 
$\chi=\pm \sqrt{\kappa ^2+\eta ^2}$ (corresponding to the solid lines in
Fig.~\ref{disp}) originate from the driven equation (\ref{driven}) and
represent the dispersion relation of a Bragg reflector with the gap
$|\chi |<\eta $, that does not feel the interaction with the active layers. 
Important roots of Eq. (\ref{dispersion}) are those of the expression in
the curly brackets, shown by the dashed and dash-dotted lines in Fig. 
\ref{figschem}. These roots correspond to nontrivial spectral features: 
bright or dark solitons in the indicated (shaded) bands.

The frequencies corresponding to $k=0$ are 
$\chi_0 = \eta$ and $\chi_{0,\pm} =-( \eta -\de )/2 \pm 
\sqrt{2+ ( \eta +\de )^2/2}$, 
while at $k^2 \to \infty$ the asymptotic expressions for
different branches of the dispersion relation are $\chi =\pm k$ and $
\chi =\delta +2\left( \eta -\delta \right) k^{-2}$. Thus, the
linearized spectrum always splits into {\em two\/} gaps, separated by
an allowed band, except for the special case, $\eta =\eta _0 \equiv
\de/2 +\sqrt{1+\de^2/4}$, when the upper gap closes
down. The upper and lower band edges are those of the periodic
structure, shifted by the induced TLA polarization in the limit of a
strong reflection. They approach the SIT spectral gap for forward- and
backward-propagating waves \cite{Mant95} in the limit of weak
reflection. The allowed middle band corresponds to a {\em polaritonic 
excitation} (collective atomic polarization) in the periodic structure.

\subsection{Standing (quiescent) self-localized pulses}

We seek the stationary solutions of Eqs. (\ref{bef1}) and (\ref{bef2})
corresponding to bright solitons in the form
\begin{equation}
\Sigma _{+}=e^{-i\chi \tau }{\cal S} (\zeta ),\qquad 
P=i\ e^{-i\chi \tau } {\cal P}(\zeta ) \label{stationary}
\end{equation}
with real ${\cal P}$ and ${\cal S}$. Substituting this into (\ref{bef2}), 
we eliminate ${\cal P}$ in favor of ${\cal S}$ and obtain an equation for 
${\cal S} (\zeta )$, 
\begin{equation}
\frac{d^2 {\cal S}}{d \zeta^2} =
(\eta ^2-\chi ^2){\cal S} -2{\cal S} \frac{(\eta -\chi
)\cdot {\rm sign}(\chi -\delta )}{\sqrt{(\chi -\delta )^2+{\cal S} ^2}}.
\label{sigma''}
\end{equation}
It then follows \cite{Kozh98a} that bright solitons can appear
in two frequency bands $\chi$, the lower band being  
$\chi _1 <\chi < {\rm min} \{\chi_2,-\eta ,\de \}$, and the upper band being
${\rm max} \{\chi _1,\eta ,\delta \}<\chi <\chi _2$, 
where $ \chi _{1,2} \equiv 1/2 [\de - \eta \mp \sqrt{(\eta +\de )^2+8}]$
are the boundary frequencies. The lower band exists for all values $\eta >0$ 
and $\delta $, while the upper one only exists for the weak-reflectivity case 
$\de > \eta -1/\eta$. On comparing these expressions with the spectrum shown 
in Fig. \ref{disp}, we conclude that part of the lower gap is always empty 
from solitons, while the upper gap is completely filled with stationary 
solitons in the weak-reflectivity case, and completely empty in the opposite 
limit. 

In an implicit form, the solution of Eq. (\ref{sigma''}) reads
\begin{equation}
{\cal S}(\zeta) = 2|\chi-\delta|{\cal R}(\zeta) \left( 1 - {\cal R}^{2}
(\zeta) \right) ^{-1}, \label{sss}
\end{equation}
with
\begin{equation}
|\zeta| = \sqrt{2\left| \frac{\chi - \delta}{\chi - \eta}  \right|}
\left[ (1-{\cal R}_{0}^{2})^{-1/2} \tan ^{-1}
\sqrt{\frac{{\cal R}_{0}^{2}-{\cal R}^{2}}
{1-{\cal R}_{0}^{2}}} + (2{\cal R}_{0})^{-1} \ln \left( \frac{{\cal R}_{0}
+ \sqrt{{\cal R}_{0}^{2}-{\cal R}^{2}}}{{\cal R}} \right)
\right] , \label{zeta}
\end{equation}
and ${\cal R}_{0}^{2} = 1-|(\chi+\eta)(\chi-\delta)|/2$. 
This zero-velocity (ZV) gap soliton is always {\em single}-humped and its
amplitude, found from Eq. (\ref{zeta}), is given by
\begin{equation}
{\cal S}_{\max }=4 {\cal R}_0/\sqrt{\left| \chi +\eta \right| }.
\label{amplitude}
\end{equation}

To calculate the electric field in the antisymmetric
$\Sigma_-$ mode, we substitute $\Sigma_-=i e^{-i\chi \theta }{\cal A} (\zeta)$
into Eq. (\ref{driven}) and obtain 
\begin{equation}
{\cal A} ^{\prime \prime }+\left( \chi ^2-\eta ^2\right) {\cal A} =
2{\cal P}^{\prime },
\label{eqalpha}
\end{equation}
which can be easily solved by the Fourier transform, once ${\cal
P}(\zeta )$ is known. We note that, depending on the parameters $\eta$,
$\delta$ and $\chi$, the main part of the soliton energy can be
carried either by the $\Sigma_{+}$ or the $\Sigma_{-}$ mode.

The most drastic difference of these new solitons from the well-known SIT 
solitons in Eq. (\ref{2pi-SG}) is that the area of the ZV soliton (integrated 
over $\zeta$) is not restricted to $2\pi$, but, instead, may take an 
{\em arbitrary} value. This basic new feature shows that the Bragg reflector 
can enhance (by {\em multiple reflections}) the field coupling to the TLA, so 
as to make the pulse area {\em effectively} equivalent to $2 \pi$. In the 
limit of the small-amplitude and small-area solitons, ${\cal R}_0^2 \ll 1$, 
Eq. (\ref{zeta}) can be easily inverted, the ZV soliton becoming a broad 
sech-like pulse:
\begin{equation}
{\cal S} \approx 2|\chi -\delta |{\cal R}_0\,{\rm sech}\left( \sqrt{2\left|
\frac{\chi -\eta }{\chi -\delta }\right| }{\cal R}_0\zeta \right) .
\label{small}
\end{equation}
In the opposite limit, $1-{\cal R}_0^2 \rightarrow 0$, i.e., for
vanishingly small $|\chi +\eta |$, the the soliton is characterized by 
a {\em broad central part} with a width 
$\sim \left( 1-{\cal R}_0^2\right) ^{-1/2}$ and its amplitude
(\ref{amplitude}) becomes very large. Thus, although the ZV soliton has a 
single hump, its shape is, in general, strongly different from that of the 
traditional nonlinear-Schr\"odinger (NLS) sech pulse.

\subsection{Moving solitons}

One could expect a translational invariance of the ZV solitons (\ref{sss})
on physical grounds. Hence, a full family of soliton solutions should
have velocity as one of its parameters. This can be explicitly
demonstrated in the limit of the small-amplitude large-width solitons
[cf. Eq. (\ref{small})]. We search for the corresponding solutions in
the form $\Sigma_+(\zeta ,\tau )={\cal S} (\zeta ,\tau )\exp (-i\chi_0\tau)$,
$P(\zeta ,\tau )=i {\cal P}(\zeta ,\tau )\exp ( -i\chi_0\tau )$ 
[cf. Eqs. (\ref{stationary})], where $\chi_0$ is the frequency corresponding 
to $k=0$ on any of the three branches of the dispersion relation 
(\ref{dispersion}) (see Fig.  \ref{disp}), and the functions 
${\cal S} (\zeta ,\tau )$ and ${\cal P}(\zeta ,\tau )$ are assumed to be 
slowly varying in comparison with $\exp ( -i\chi_0\tau )$. Under these 
assumptions, we arrive at the following asymptotic equation for 
${\cal S} (\zeta ,\tau )$:
\begin{equation}
\left[ 2i\frac{\chi_0(\chi_0-\de )^2-\eta +\de}{(\chi_0-\de)^2} 
\frac{\partial}{\partial \tau } + \frac{\partial^2}{\partial \zeta ^2}
+\frac{\chi_0-\eta }{(\chi_0-\de)^3} |{\cal S}|^2 \right] {\cal S} = 
\left( \eta ^2-\chi_0^2 + 2 \frac{\chi_0-\eta }{\chi_0-\de} \right) {\cal S} .
\label{NLS}
\end{equation}
Since this equation is of the NLS form, it has the full two-parameter
family of soliton solutions, including the moving ones \cite{Newe92}.

\begin{figure}[t]
\centerline{\includegraphics[height=6cm]{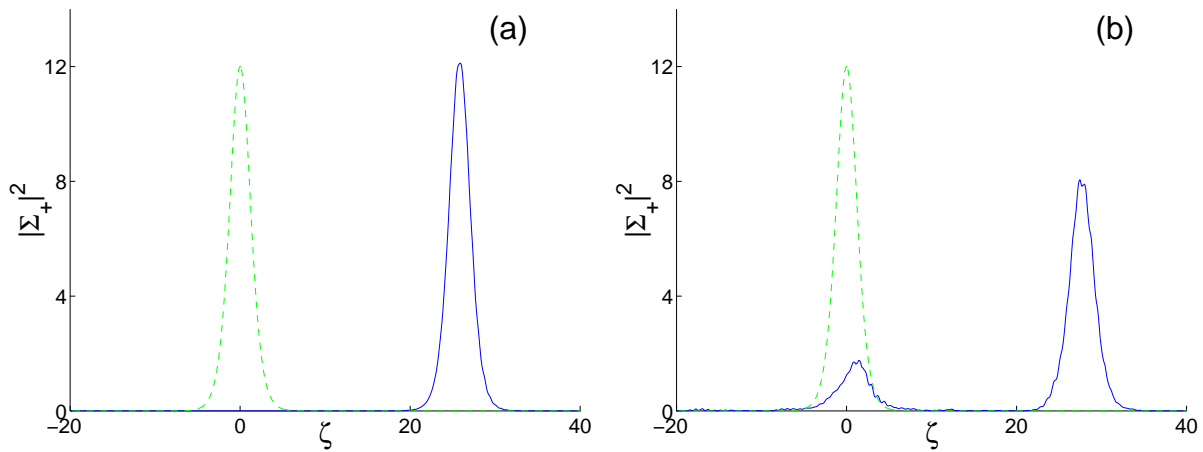}}
\caption{Pulses obtained as a result of `pushing' a zero-velocity RABR 
soliton (dashed lines): (a) push, characterized by the initial multiplier
$\exp (-ip\zeta )$ after a sufficiently long evolution ($\tau=400$)
(solid lines). $\delta=0$, $\eta=4$, $\chi=-4.4$, and $p=0.1$. (b)
idem, but for $p=0.5$. 
\label{fig:push}}
\end{figure}

In order to check the existence and stability of the moving
solitons numerically, the following procedure has been used \cite{Kozh98a}: 
Eqs. (\ref{bef1}) and (\ref{bef2}) were simulated for an
initial configuration in the form of the ZV soliton multiplied by
$\exp (ip\zeta )$ with some wavenumber $p$, in order to `push' the
soliton. The results demonstrate that, at sufficiently small $p$, the
`push' indeed produces a moving stable soliton [Fig. \ref{fig:push}(a)]. 
However, if $p$ is large enough, the multiplication by $\exp (ip\zeta )$ 
turns out to be a more violent perturbation, splitting the initial pulse 
into two solitons, one quiescent and one moving [Fig. \ref{fig:push}(b)].

\subsection{Light bullets (spatiotemporal solitons) in PCs 
\label{sec:lightbullets}}

The advantageous properties of SIT GS can be supplemented by immunity to 
transverse diffraction, i.e., {\em simultaneous} transverse and longitudinal 
self-localization of light in a PC: multi-dimensional spatio-temporal solitons 
or ``light bullets'' (LBs) \cite{Silb90} have been analytically and 
numerically predicted by our group to exist and be stable, not only in 
uniform 2D and 3D SIT media \cite{Blaa00}, but also in  2D or 3D periodic 
structures, wherein SIT solutions combining LB and GS properties are 
demonstrated \cite{Blaa00a}. Our objective is to consider the propagation 
of an electromagnetic wave with a frequency close to $\om_c$ through a 2D 
PC doped by thin TLA layers. The forward- and backward-propagating components 
satisfy equations that are a straightforward generalization of the 1D 
equations (\ref{bef1}) and (\ref{driven})
\begin{eqnarray}
-i \frac{\partial^3 \Sigma_+}{\partial \tau x^2} + 
i \frac{\partial^3 \Sigma_-}{\partial \zeta x^2} + 
\frac{\partial^2 \Sigma_+}{\partial \tau^2} - 
\frac{\partial^2 \Sigma_+}{\partial \zeta^2} +
\eta \frac{\partial^2 \Sigma_+}{\partial x^2} + \eta^{2}
\Sigma_{+}-2\frac{\partial P}{\partial \tau }-2i\eta {P} 
= 0, \label{lb1} \\ 
-i\frac{\partial^3 \Sigma_-}{\partial \tau x^2} + i 
\frac{\partial^3 \Sigma_+}{\partial \zeta x^2} + 
\frac{\partial^2 \Sigma_-}{\partial \tau^2} -
\frac{\partial^2 \Sigma_-}{\partial \zeta^2} -
\eta \frac{\partial^2 \Sigma_-}{\partial x^2} + \eta^{2}
\Sigma_{-}+2\frac{\partial P}{\partial \zeta} = 0, \label{lb2}   
\end{eqnarray}
where the Fresnel number $F>0$, which governs the transverse diffraction in 
the 2D and 3D propagation, was incorporated into $x$ denoting the transverse
coordinate. The equations for the polarization $P$ and inversion $w$ are the 
same as Eqs. (\ref{P}) and (\ref{w}). 

We search for analytical LB solutions of Eqs. (\ref{lb1}), (\ref{lb2}), 
(\ref{P}) and (\ref{w}), by the following ansatz that reduces in 1D to the 
exact moving GS \cite{Kuri2001,Kozh95}   
\begin{eqnarray}
\Sigma_+ &=& A_0 \sqrt{\text{sech} \Theta_1 \text{sech} \Theta_2} 
e^{i(\kappa \zeta - \chi \tau)+i\pi /4}, \label{lbs1} \\
\Sigma_- &=& \Sigma_+/v, \label{lbs2} \\
P &=& \sqrt{ \text{sech} \Theta_1 \text{sech} \Theta_2}
\{(\tanh \Theta_1+\tanh \Theta_2)^{2}+ \nonumber \\
& & { } \frac{\de - \eta}{4\eta} C^{4} 
\bigl[ (\tanh \Theta_1-\tanh \Theta_2)^{2}-2(\text{sech}^{2} \Theta_1 
+ \text{sech}^2 \Theta_2)\bigr]^2\}^{1/2} 
e^{i(\kappa \zeta - \chi \tau)+i\nu }, \label{lbs3} \\
w &=&\left[ 1-|{P}|^{2}\right] ^{1/2}, \label{lbs4}
\end{eqnarray}
with $\Theta_1 (\tau , \zeta) \equiv \beta ( \zeta - v \tau ) +
\Theta_0 + C x$, $\Theta_2(\tau ,\zeta) \equiv \beta ( \zeta - v \tau ) + 
\Theta_0-Cx$, the phase $\nu$ and coefficients $\Theta_0$ and $C$ being 
real constants, while the other parameters are defined as
$ A_0 =  2 \sqrt{\de /\eta -1}$, $\beta = \sqrt{\de /\eta +1}$, 
$v =- \sqrt{(\de - \eta)/(\de + \eta)}$, $\kappa = -\sqrt{\de^2 - \eta^2}$, 
and $\chi=\de$. 

\begin{figure}[t]
\centerline{\includegraphics[height=6cm]{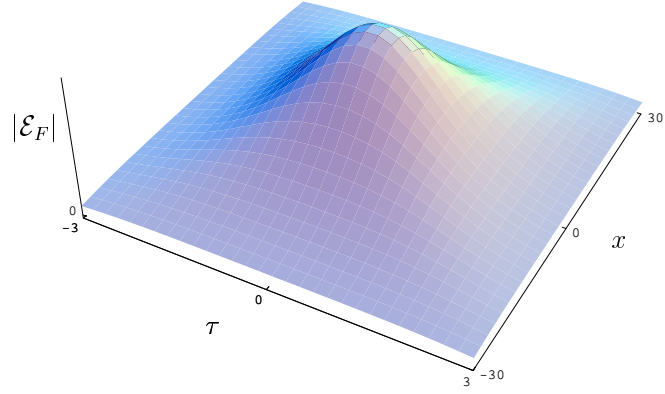}}
\caption{The forward-propagating electric field of the two-dimensional 
`light bullet' in the Bragg reflector, $|{\cal E}_F|$, vs. time $\tau $ 
and transverse coordinate $x$, after having propagated the distance 
$z=1000$. The parameters are $\eta = 0.1$, $\delta = 0.2$, $C=0.1$ and 
$\Theta _{0}=-1000$. The field is scaled by the constant 
$\hbar /4\tau _{0}\mu n_{0}$.  
\label{Bullets}}
\end{figure}

The ansatz (\ref{lbs1})- (\ref{lbs4}) satisfies Eqs. (\ref{lb1}) and 
(\ref{lb2}) exactly, while Eqs. (\ref{P}) and (\ref{w}) are satisfied to
order $\sqrt{\de/\eta-1}C^{2}$, which requires that 
$\sqrt{\de/\eta-1} C^2 \ll 1$. The ansatz applies for {\it arbitrary} $\eta $, 
admitting {\em both} weak ($\eta \ll 1$) and strong ($\eta >1$) reflectivities 
of the Bragg grating, provided that the detuning remains small with respect 
to the gap frequency. Comparison with numerical simulations of 
Eqs. (\ref{lb1}), (\ref{lb2}), (\ref{P}) and (\ref{w}), using 
Eqs. (\ref{lbs1})-(\ref{lbs4}) as an initial configuration, tests this 
analytical approximation and shows that it is indeed fairly close to a 
numerically exact solution; in particular, the shape of the bullet remains 
within 98\% of its originally presumed shape after having propagated a 
large distance, as is shown in Fig. \ref{Bullets}.

Three-dimensional (3D) LB solutions with axial symmetry have also been
constructed in an approximate analytical form and successfully tested
in direct simulations, following a similar approach \cite{Blaa00a}.
Generally, they are not drastically different from their 2D
counterparts described above.

\subsection{Information transmission by SIT GSs and LBs}

The efficiency of information transmission is characterized either by 
channel (information) capacity $C = W \ln (I_s /I_n)$, where $W$ is the 
bandwidth and $I_s/I_n$ is the ratio of the signal-to-noise intensities, or
by the data transmission density $D=N M$, where $N$ is the number of
bits per channel and $M$ is the number of accessible channels. Both $C$ and $D$
can be very high in the case of a SIT GS or LB for the following reasons:
(a) The bandwidth $W$ is large, being limited by the PBG width, which can
be very large in the optical domain. At the same time, {\em noise is very 
effectively suppressed} by the Bragg reflection and by the absence of 
diffraction losses in the case of a LB. 
(b) The maximal transmission density
$D$ can be estimated \cite{Mossberg} as the ratio of the accessible bandwidth,
in our case the PBG width (in excess of $10^{13}$ s$^{-1}$ in the optical 
domain), to the spontaneous linewidth ($10^{6}$ s$^{-1}$ for rare-earth ions).
Hence, these modes of transmission can be very effective for optical 
communications.

\section{Cross coupling between electromagnetically-induced and 
self-induced transparency pulses
\label{sec:eit_sit}}

\subsection{EIT in bulk media: Background
\label{eit_bg}}

\begin{figure}[t]
\centerline{\includegraphics[height=6cm]{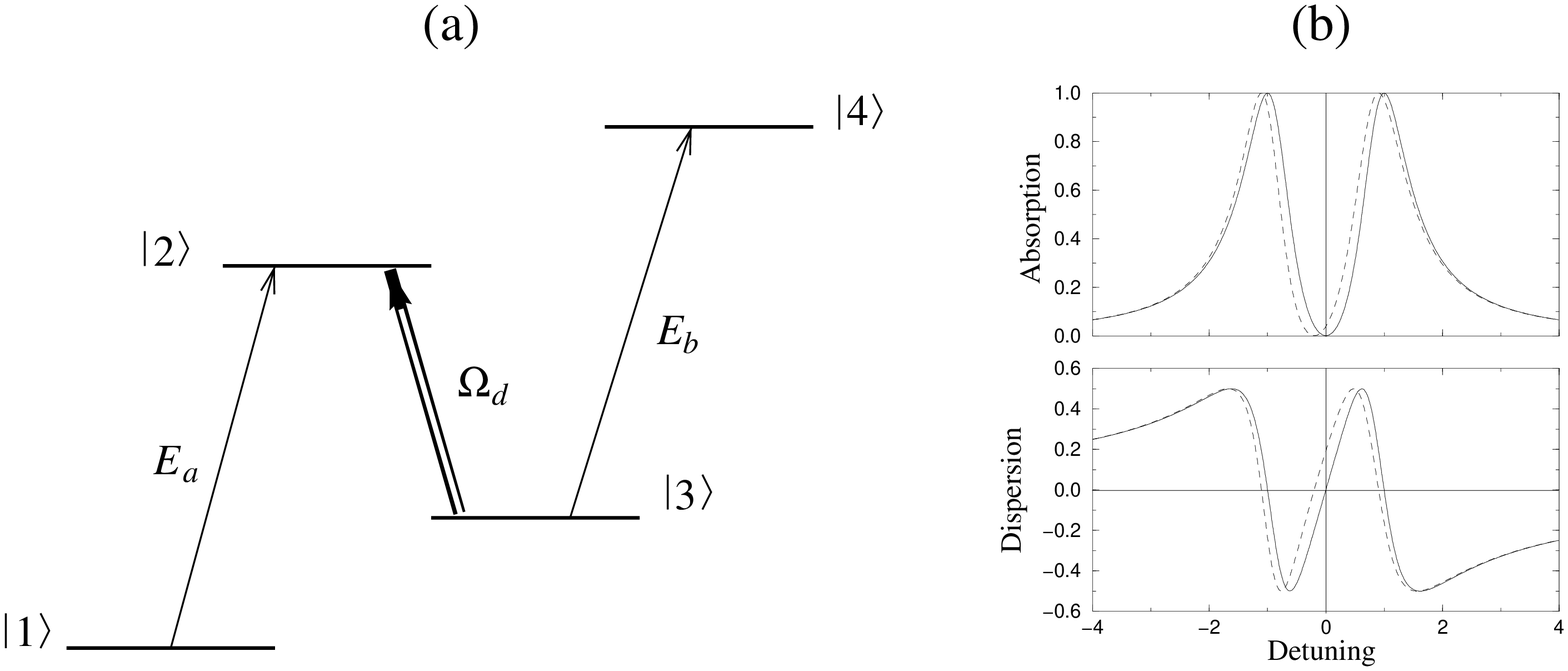}}
\caption{(a) Schematic representation of the atomic system interaction with 
strong driving field on the transition $\ket{2}\to \ket{3}$ and weak fields 
$E_a$ and $E_b$ on the transitions $\ket{1}\to \ket{2}$ and 
$\ket{3}\to \ket{4}$, respectively. 
(b) Absorption and dispersion spectra of the $E_a$ field in the absence (solid 
line) or presence (dashed line) of the $E_b$ field.
\label{imam}}
\end{figure}

Electromagnetically induced transparency (EIT) is based on the phenomenon of 
coherent population trapping \cite{eit_a,eit_b}, in which the application of 
two laser fields to a three-level atomic system creates the so-called 
``dark state'', which is stable against absorption of both fields. Consider 
a four-level atomic system whose level configuration is depicted in 
Fig. \ref{imam}(a). The phase shift and absorption of an optical field $E_i$ 
are given by the real and imaginary parts of its complex polarizability 
$\alpha_i$. In the absence of the ``control'' field $E_b$, the usual EIT 
spectrum [Fig. \ref{imam}(b)] for the weak probe field $E_a$ exhibits 
vanishing phase shift and absorption 
[$\text{Re} (\alpha_a) = \text{Im} (\alpha_a) = 0$] at the two-photon Raman 
resonance $\om_a = \om_{21}+\om_d$, where $\om_a$ and $\om_d$ are the 
frequencies of the probe and driving fields, respectively, and $\om_{ij}$ is 
the frequency of the  atomic transition $\ket{i}\to \ket{j}$. An off-resonant 
control field $E_b$ with the frequency $\om_b$ such that 
$|\De_b| = |\om_b - \om_{43}| \gg \ga_4$, where $\ga_i$ is the decay rate of 
the corresponding atomic level, induces an ac Stark shift of level $\ket{3}$ 
and thereby shifts the EIT spectrum [Fig. \ref{imam}(b)]. 

Due to the steepness of the dispersion curve in the vicinity of the Raman 
resonance, $|\partial_{\om_a} \text{Re} (\alpha_a)| \gg 
|\partial_{\om_a} \text{Im} (\alpha_b)|$, this Stark shift leads to a large 
phase shift along with small absorption of the probe field $E_a$: 
\begin{equation}
\phi_a = \text{Re} (\alpha_a) z \simeq 
-\frac{\alpha_0 \ga_2 |\Om_b|^2}{2 \De_b |\Om_d|^2} z, \;\;\;
\text{Im}(\alpha_a)= - \frac{\ga_4 \text{Re}(\alpha_a)}{2\De_b} \ll 
\text{Re} (\alpha_a),
\end{equation}
where $\alpha_0$ is the resonant absorption coefficient of the medium at the 
frequency $\om_{21}$ and $\Om_i = \mu_{ij} E_i/\hbar$ is the Rabi frequency of 
the corresponding field ($\mu_{ij}$ the dipole matrix element on the 
respective transition). This is the essence of the so-called {\em giant Kerr 
cross-phase modulation} of a probe field by a control field, introduced first 
by Schmidt and Imamo\u{g}lu \cite{imam}. Later Harris and Yamamoto 
\cite{harris_a} have predicted that a resonant control field $E_b$ with 
$|\De_b| < \ga_4$ can destroy the coherence between the two ground levels 
$\ket{1}$ and $\ket{3}$, which leads to a two-photon absorption 
$\text{Im} (\alpha_{a,b}) = 
\frac{\alpha_0 \ga_2|\Om_{b,a}|^2}{\ga_4 |\Om_d|^2}$,
that is, the medium absorbs two fields simultaneously, but does not absorb 
one field alone. This is the essence of a probe-photon switch, conditional
on the presence of control photons.

The main limitation of the above schemes \cite{imam,harris_a,harris_b} stems 
for the fact that the {\it effective interaction length is limited} by the 
mismatch between the group velocity of the slowly propagating $E_a$ field, 
$v^{(a)}_g \simeq \frac{2 |\Om_d|^2}{\alpha_0 \ga_3} \ll c/n_0$ and that of the
nearly-free propagating $E_b$ field, $v^{(b)}_g \simeq c/n_0$. For weak 
(few-photon) pulses, this mismatch ultimately limits the maximal phase shift
or absorption of the probe in the presence of the control field.

\subsection{Simultaneous EIT and SIT in RABR 
\label{eit_sit}}

\begin{figure}[t]
\centerline{\includegraphics[height=5.5cm]{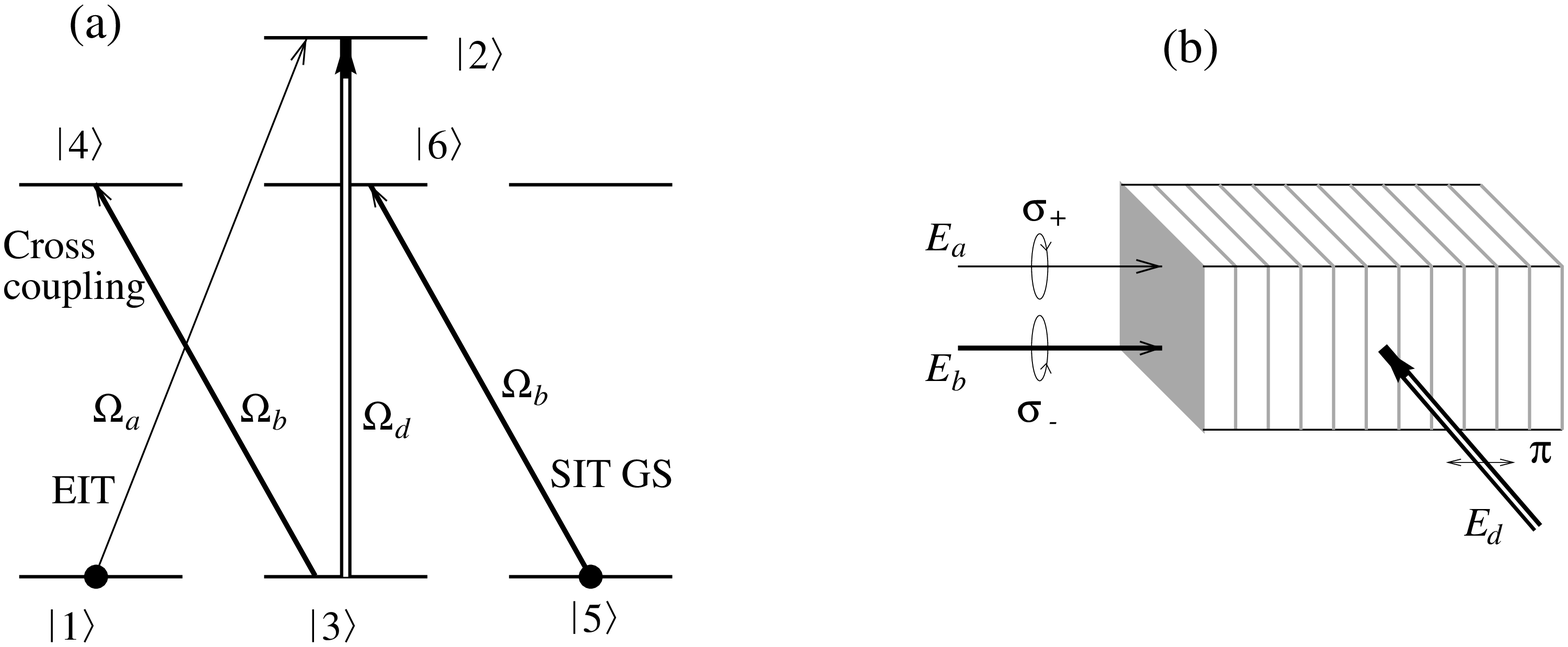}}
\caption{(a) Schematic representation of the field-atom system:
The probe field $E_a$ exhibits EIT on the transition $\ket{1} \to \ket{2}$,
in the presence of the driving field $E_d$ on the transition 
$\ket{3} \to \ket{2}$; The control field $E_b$ exhibits SIT GS on the 
transitions $\ket{5} \to \ket{6}$; The transition $\ket{3} \to \ket{4}$
serves to cross-couple the two fields. Initially only the states $\ket{1}$ 
and $\ket{5}$ are populated.
(b) Polarizations and propagation directions of the fields involved in the
transitions above.
\label{eit-sit}}
\end{figure}

In this section we propose a new implementation of the cross-phase 
modulation, in which the {\em group velocities of both fields can be matched}, 
allowing one to obtain any desired phase shift of the probe field with a 
weak control field. To this end, we consider the same configuration as 
in Sec. {\ref{sit_rabr}}, leading to SIT GS and LB solutions: a PC periodically
doped by thin layers of atoms at the maxima of its refractive index. However,
the multi-level structure of the atoms is now playing a role: it is shown in
Fig. \ref{eit-sit}, along with the polarizations and propagation directions of
the fields involved. The states $\ket{1}$, $\ket{3}$ and $\ket{5}$ are the 
degenerate Zeeman components with $M_F=-1,0,+1$, respectively, of the atomic 
ground level having total angular momentum $F=1$. Similarly, the states 
$\ket{4}$ and  $\ket{6}$ are the degenerate Zeeman components with $M_F=-1,0$, 
respectively, of the excited level having angular momentum $F=1$. Finally, 
the state $\ket{2}$ corresponds to the single Zeeman component with $M_F=0$ 
of another excited level having $F=0$. Such a level scheme is found, e.g., in 
alkali atoms, where the ground level is $S_{1/2},F=1$ and the two excited 
levels are $P_{1/2},F=1$ and $P_{3/2},F=0$. Due to the dipole selection rules, 
the $\pi$-polarized driving field couples the states with $\De M=0$, the 
$\sigma_+$-polarized $E_a$ field couples the states with $\De M=1$ and the 
$\sigma_-$-polarized $E_b$ field couples the states with $\De M=-1$. 

We assume that initially all the atoms are optically pumped into the states 
$\ket{1}$ and $\ket{5}$, which then acquire equal populations 1/2. Hence, the 
sequence of transitions $\ket{1} \to \ket{2} \to \ket{3} \to \ket{4}$ repeats 
that of Fig. \ref{imam}(a), realizing the cross-phase modulation scheme of
Sec. {\ref{sec:eit_sit}\ref{eit_bg}}. The frequency of the $E_a$ field is 
far from the band gap frequencies of the PC, while the frequency of the $E_b$ 
field is within the band gap. 

As was shown in Sec. {\ref{sit_rabr}}, PC structures doped with the 
near-resonant TLAs can support standing and slowly moving SIT GSs, whose 
pulse area (integrated over $z$) can take an {\em arbitrarily small} value. 
In the present setup, the transition $\ket{5} \to \ket{6}$ realizes that 
near-resonant TLA, allowing for slow propagation of the $E_b$ field through 
the PC. 

Let us write the propagation equation for the slowly moving SIT soliton in 
the form
\begin{equation}
\Om_{\texts{SIT}} (z,t) = \Om_b \text{sech} 
\left( \frac{t v_g^{(b)} - z}{2 \beta }\right), \label{rabi_sit}
\end{equation}
where $\Om_b$ is the peak Rabi frequency, $v_g^{(b)}$ is the group velocity
and $\beta = (2 \alpha_b)^{-1}$, with $\alpha_b$ being the absorption 
coefficient of the active medium at the carrier frequency $\om_b$ of the 
soliton. The temporal width of the pulse is given by 
$\tau_b = 2 \beta /v_g^{(b)}$. The area of the $E_b$ pulse
\begin{equation}
\theta_b = \int_{-\infty}^{\infty} \Om_{\texts{SIT}} d t = 
\frac{\Om_b}{v_g^{(b)} \alpha_b} \pi,
\end{equation}
is then inversely proportional to the group velocity of the pulse. Hence, 
the SIT condition (Sec. {\ref{sec:eit_sit}\ref{eit_bg}}) $\theta_b = 2 \pi$ 
imposes a unique relation between the Rabi frequency of the SIT soliton and 
its group velocity:
\begin{equation} 
v_g^{(b)} = \frac{\Om_b}{2\alpha_b} .
\end{equation} 
The absorption-free propagation of the SIT soliton is limited to 
$z< v_g^{(b)}/\ga_6$, where $\ga_6$ is the decay rate of the upper atomic
state $\ket{6}$.

Our aim is to match the group velocities of the $E_a$ field subject to EIT
and the $E_b$ field having the form of a slow SIT gap soliton: 
$v_g^{(a)} = v_g^{(b)}$. This requires that 
$|\Om_d|^2 = \Om_b \alpha_a \ga_2 /4\alpha_b$, i.e., an appropriate choice of 
the driving field Rabi frequency $\Om_d$, for a given Rabi frequency $\Om_b$ 
of the control field. Such velocity matching of the two copropagating weak 
fields would {\em maximize their interaction}. 

One possibility to launch the required slow SIT soliton is to irradiate the PC 
by a laser beam at a small angle $\psi$ relative to the periodicity direction 
$z$, $\psi \simeq D n_0 v_g^{(b)} /(L c) \ll 1$, where $D$ and $L$ are, 
respectively, the transverse and longitudinal dimensions of the structure. 
This choice of $\psi$ ensures that the $z$-component of the beam, which 
forms the SIT soliton and propagates in the PC with the group velocity 
$v_g^{(b)}$ over the distance $L$, will traverse the structure during 
the same time as the transverse component of that beam, which covers the 
distance $D$ with the velocity $(c/n_0)\sin \psi$.

We have checked that for the parameter values corresponding to dopant atoms
(or ions) with the mean density $N=10^{13}$ cm$^{-3}$ (surface density 
of $4 \times 10^8$ cm$^{-2}$ in the thin layers), $\De_b =30 \ga_4$, 
$|\Om_b| \simeq 10^6$ rad/s and $|\Om_d| \simeq 4\times 10^6$ rad/s, we obtain
$\pi$ phase shift of the $E_a$ field over a distance $z \simeq 4$ cm, while
the absorption probability remains less that 10\%.

\begin{figure}[t]
\centerline{\includegraphics[height=5.8cm]{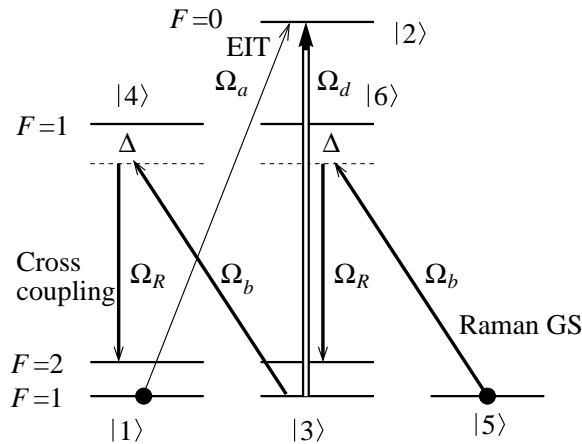}}
\caption{Schematic representation of the level scheme suitable for EIT
in the presence of a Raman GS (only the relevant levels are shown)
\label{ram-sit}}
\end{figure}

One possible difficulty of our scheme is that, with the parameters listed 
above, the temporal width of the $E_b$ field is $\tau_b \sim 10^{-6}$ s, and 
the interaction time is of the order of $10^{-3}$ s, while the lifetime of the 
SIT soliton is of the order of the decay time of the excited atomic state 
$\ga_6^{-1} \sim 10^{-7}$ s. One can cope with this problem by employing the 
atomic level scheme shown in Fig. \ref{ram-sit}, which allows one to launch 
{\em Raman solitons}. We irradiate the system with an additional 
strong cw field $E_R$, which couples the $F=1$ excited state with the
$F=2$ metastable ground states. The fields $E_R$ and $E_b$ are largely
detuned from the fast decaying excited states $\ket{4}$ and $\ket{6}$ by
an amount $\De \gg \ga_{4,6}$. Then, upon adiabatically eliminating the
states $\ket{4}$ and $\ket{6}$, we obtain that the Rabi frequency $\Om_b$ 
of the control field in Eq. (\ref{rabi_sit}) is simply replaced by 
$\Om_b \Om_R/\De$. The lifetime of the SIT soliton is given now
by the lifetime of the $F=2$ ground states, which can be very large, reaching
in some instances a fraction of a second! In addition, such a setup allows
one to launch slow Raman GSs \cite{perlin}, and thus circumvent the difficulty
of launching a standing (ZV) or slowly moving GSs, which must otherwise
overcome the high reflectivity of the PC boundaries.

\section{Conclusions}
\label{sec:concl}

In this paper we have focused on properties of solitons in a doped PC or RABR, 
combining a periodic refractive-index superlattice (Bragg reflector in 1D or 
2D) and a periodic set of thin active layers (consisting of TLAs 
{\em resonantly} interacting with the field). We have demonstrated that the 
system supports a vast family of bright GSs, whose properties differ 
substantially from their counterparts in periodic structures with either cubic 
or quadratic {\em off-resonant} nonlinearities. Depending on the initial 
conditions, these can be either standing (ZV) or slowly moving stable solitons 
that exhibit SIT irrespective of their photon number (pulse energy) for an
appropriate group velocity.
A multidimensional version of this model corresponds to a periodic 
set of thin active layers placed at the maxima of a 2D- or 3D-periodic 
refractive index. It has been found to support stable propagation of 
spatiotemporal solitons in the form of 2D- and 3D-localized LBs.

The best prospect of realizing a PC which is adequate for observing
the GSs and LBs is to use thin layers of {\em rare-earth ions} 
\cite{Grei99} embedded in a spatially-periodic semiconductor structure 
\cite{Khit99}.  The TLAs in the layers should be rare-earth-ions 
with the density of $10^{15}-10^{16}$ cm$^{-3}$, and large transition dipole 
moments. The parameter $\eta $ can vary from 0 to 100 and the detuning is 
$\sim 10^{12}-10^{13}$ s$^{-1}$.
Cryogenic conditions in such structures can strongly extend the dephasing 
time $T_2$ and thus the soliton's or LB's lifetime, well into the $\mu$sec 
range \cite{Grei99}, which would greatly facilitate the experiment. 

In a 2D PC, LBs can be envisaged to be localized on the time and 
transverse-length scales, respectively, $\sim 10^{-12}$ s and $1\,\mu $m.  
The incident pulse has uniform transverse intensity and the transverse 
diffraction is strong enough. One needs $d^{2}/l_{\rm abs}\lambda _{0}<1$, 
where $l_{{\rm abs}}$, $\lambda_0$ and $d$ are the resonant-absorption
length, carrier wavelength, and the pulse diameter, respectively.
For $l_{{\rm abs}}\sim 10^{-3}$\thinspace\ m and $\lambda _{0}\sim
10^{-4}$\thinspace\ m, one thus requires $d<10^{-4}$ m, which implies
that the transverse size of the PC must be a few $\mu$m.

We have considered here (Sec. {\ref{sec:eit_sit}\ref{eit_sit}}) the 
cross-coupling of optical beams in a PC or RABR. We have pointed out, for 
the first time, the advantageous features of the cross coupling between 
EIT and SIT pulses, which is capable of producing extremely strong 
correlations between the two pulses. With doping parameters as above, 
and driving and control fields with Rabi frequencies of the order of 
$10^6$ rad/s, we can obtain a phase shift of $\pi$ for the weak probe 
pulse over a distance of few cm. This is much larger than any corresponding 
phase shift (for similar control fields) in other media.

We strongly believe that the highly promising payoff expected from the
construction of suitable structures justifies the experimental challenge they 
pose. If and when the schemes proposed above are experimentally realized, 
they may prove to be useful for producing ultrasensitive nonlinear 
phase shifters or logical photon switches for both classical and quantum 
information processing or communication, owing to the unique advantages of 
the doped PCs over conventional EIT schemes \cite{imam,harris_a,harris_b,lukin}
or high-Q cavities \cite{cavity_a,cavity_b}:

\section*{Acknowledgments}

This work was supported by the EU (ATESIT) Network, the US-Israel BSF and the 
Feinberg Fellowship (D.P.).


\end{document}